

\raggedbottom

\def\refto#1{$^{#1}$}           
\def\ref#1{ref.~#1}                     
\def\Ref#1{#1}                          
\gdef\refis#1{\item{#1.\ }}                    
\def\beginparmode{\endmode
  \begingroup \def\endmode{\par\endgroup}}
\let\endmode=\par
\def\body{\beginparmode}
\def\head#1{                    
  \goodbreak\vskip 0.5truein    
  {\centerline{\bf{#1}}\par}
   \nobreak\vskip 0.25truein\nobreak}
\def\references                 
Phys Rev
  {\head{References}            
co
mmas)\.
   \beginparmode
   \frenchspacing \parindent=0pt \leftskip=1truecm
   \parskip=8pt plus 3pt \everypar{\hangindent=\parindent}}
\def\endreferences{\body}

\catcode`@=11
\newcount\r@fcount \r@fcount=0
\newcount\r@fcurr
\immediate\newwrite\reffile
\newif\ifr@ffile\r@ffilefalse
\def\w@rnwrite#1{\ifr@ffile\immediate\write\reffile{#1}\fi\message{#1}}

\def\writer@f#1>>{}
\def\referencefile{
  \r@ffiletrue\immediate\openout\reffile=\jobname.ref%
  \def\writer@f##1>>{\ifr@ffile\immediate\write\reffile%
    {\noexpand\refis{##1} = \csname r@fnum##1\endcsname = %
     \expandafter\expandafter\expandafter\strip@t\expandafter%
     \meaning\csname r@ftext\csname r@fnum##1\endcsname\endcsname}\fi}%
  \def\strip@t##1>>{}}

\def\citeall#1{\xdef#1##1{#1{\noexpand\cite{##1}}}}
\def\cite#1{\each@rg\citer@nge{#1}}	

\def\each@rg#1#2{{\let\thecsname=#1\expandafter\first@rg#2,\end,}}
\def\first@rg#1,{\thecsname{#1}\apply@rg}	
\def\apply@rg#1,{\ifx\end#1\let\next=\relax
\else,\thecsname{#1}\let\next=\apply@rg\fi\next}

\def\citer@nge#1{\citedor@nge#1-\end-}	
\def\citer@ngeat#1\end-{#1}
\def\citedor@nge#1-#2-{\ifx\end#2\r@featspace#1 
  \else\citel@@p{#1}{#2}\citer@ngeat\fi}	
\def\citel@@p#1#2{\ifnum#1>#2{\errmessage{Reference range #1-#2\space is bad.}%
    \errhelp{If you cite a series of references by the notation M-N, then M and
    N must be integers, and N must be greater than or equal to M.}}\else%
 {\count0=#1\count1=#2\advance\count1 by1\relax\expandafter\r@fcite\the\count0,%
  \loop\advance\count0 by1\relax
    \ifnum\count0<\count1,\expandafter\r@fcite\the\count0,%
  \repeat}\fi}

\def\r@featspace#1#2 {\r@fcite#1#2,}	
\def\r@fcite#1,{\ifuncit@d{#1}
    \newr@f{#1}%
    \expandafter\gdef\csname r@ftext\number\r@fcount\endcsname%
                     {\message{Reference #1 to be supplied.}%
                      \writer@f#1>>#1 to be supplied.\par}%
 \fi%
 \csname r@fnum#1\endcsname}
\def\ifuncit@d#1{\expandafter\ifx\csname r@fnum#1\endcsname\relax}%
\def\newr@f#1{\global\advance\r@fcount by1%
    \expandafter\xdef\csname r@fnum#1\endcsname{\number\r@fcount}}

\let\r@fis=\refis			
\def\refis#1#2#3\par{\ifuncit@d{#1}
   \newr@f{#1}%
   \w@rnwrite{Reference #1=\number\r@fcount\space is not cited up to now.}\fi%
  \expandafter\gdef\csname r@ftext\csname r@fnum#1\endcsname\endcsname%
  {\writer@f#1>>#2#3\par}}

\def\ignoreuncited{
   \def\refis##1##2##3\par{\ifuncit@d{##1}%
     \else\expandafter\gdef\csname r@ftext\csname r@fnum##1\endcsname\endcsname%
     {\writer@f##1>>##2##3\par}\fi}}

\def\r@ferr{\endreferences\errmessage{I was expecting to see
\noexpand\endreferences before now;  I have inserted it here.}}
\let\r@ferences=\references
\def\references{\r@ferences\def\endmode{\r@ferr\par\endgroup}}

\let\endr@ferences=\endreferences
\def\endreferences{\r@fcurr=0
  {\loop\ifnum\r@fcurr<\r@fcount
    \advance\r@fcurr by 1\relax\expandafter\r@fis\expandafter{\number\r@fcurr}%
    \csname r@ftext\number\r@fcurr\endcsname%
  \repeat}\gdef\r@ferr{}\endr@ferences}


\let\r@fend=\endpaper\gdef\endpaper{\ifr@ffile
\immediate\write16{Cross References written on []\jobname.REF.}\fi\r@fend}

\catcode`@=12

\citeall\refto		
\citeall\ref		%
\citeall\Ref		%

\def\singlespace{\baselineskip 12pt \lineskip 1pt \parskip 2pt plus 1 pt}

\def\today{\number\day\enspace
     \ifcase\month\or January\or Febuary\or March\or April\or May\or
     June\or July\or August\or September\or October\or
     November\or December\fi \enspace\number\year}
\def\clock{\count0=\time \divide\count0 by 60
    \count1=\count0 \multiply\count1 by -60 \advance\count1 by \time
    \number\count0:\ifnum\count1<10{0\number\count1}\else\number\count1\fi}
\footline={\hss -- \folio\ -- \hss}

\def\deg{\ifmmode^\circ\else$^\circ$\fi}
\def\solar{\ifmmode_{\mathord\odot}\else$_{\mathord\odot}$\fi}
\def\jref#1 #2 #3 #4 {{\par\noindent \hangindent=3em \hangafter=1 
      \advance \rightskip by 5em #1, {\it#2}, {\bf#3}, #4.\par}}
\def\ref#1{{\par\noindent \hangindent=3em \hangafter=1 
      \advance \rightskip by 5em #1.\par}}
\newcount\eqnum
\def\nexteq{\global\advance\eqnum by1 \eqno(\number\eqnum)}
\def\lasteq#1{\if)#1[\number\eqnum]\else(\number\eqnum)\fi#1}
\def\preveq#1#2{{\advance\eqnum by-#1
    \if)#2[\number\eqnum]\else(\number\eqnum)\fi}#2}
\def\endtable{\endgroup}
\def\tableheight{\vrule width 0pt height 8.5pt depth 3.5pt}
{\catcode`|=\active \catcode`&=\active 
    \gdef\tabledelim{\catcode`|=\active \let|=\vbar
                     \catcode`&=\active \let&=\nobar} }
\def\table{\begingroup
    \def\twidth{\hsize}
    \def\tablewidth##1{\def\twidth{##1}}
    \def\defaultheight{\vrule width 0pt height 8.5pt depth 3.5pt}
    \def\heightdepth##1{\dimen0=##1
        \ifdim\dimen0>5pt 
            \divide\dimen0 by 2 \advance\dimen0 by 2.5pt
            \dimen1=\dimen0 \advance\dimen1 by -5pt
            \vrule width 0pt height \the\dimen0  depth \the\dimen1
        \else  \divide\dimen0 by 2
            \vrule width 0pt height \the\dimen0  depth \the\dimen0 \fi}
    \def\spacing##1{\def\defaultheight{\heightdepth{##1}}}
    \def\nextheight##1{\noalign{\gdef\tableheight{\heightdepth{##1}}}}
    \def\end{\cr\noalign{\gdef\tableheight{\defaultheight}}}
    \def\zerowidth##1{\omit\hidewidth ##1 \hidewidth}    
    \def\hline{\noalign{\hrule}}
    \def\skip##1{\noalign{\vskip##1}}
    \def\bskip##1{\noalign{\hbox to \twidth{\vrule height##1 depth 0pt \hfil
        \vrule height##1 depth 0pt}}}
    \def\header##1{\noalign{\hbox to \twidth{\hfil ##1 \unskip\hfil}}}
    \def\bheader##1{\noalign{\hbox to \twidth{\vrule\hfil ##1 
        \unskip\hfil\vrule}}}
    \def\spanloop{\span\omit \advance\mscount by -1}
    \def\extend##1##2{\omit
        \mscount=##1 \multiply\mscount by 2 \advance\mscount by -1
        \loop\ifnum\mscount>1 \spanloop\repeat \ \hfil ##2 \unskip\hfil}
    \def\vbar{&\vrule&}
    \def\nobar{&&}
    \def\hdash##1{ \noalign{ \relax \gdef\tableheight{\heightdepth{0pt}}
        \toks0={} \count0=1 \count1=0 \putout##1\end 
        \toks0=\expandafter{\the\toks0 &\end} \xdef\piggy{\the\toks0} }
        \piggy}
    \let\e=\expandafter
    \def\putspace{\ifnum\count0>1 \advance\count0 by -1
        \toks0=\e\e\e{\the\e\toks0\e&\e\multispan\e{\the\count0}\hfill} 
        \fi \count0=0 }
    \def\putrule{\ifnum\count1>0 \advance\count1 by 1
        \toks0=\e\e\e{\the\e\toks0\e&\e\multispan\e{\the\count1}\leaders\hrule\hfill}
        \fi \count1=0 }
    \def\putout##1{\ifx##1\end \putspace \putrule \let\next=\relax 
        \else \let\next=\putout
            \ifx##1- \advance\count1 by 2 \putspace
            \else    \advance\count0 by 2 \putrule \fi \fi \next}   }
\def\tablespec#1{
    \def\vdimens{\noexpand\tableheight}
    \def\tabby{\tabskip=0pt plus100pt minus100pt}
    \def\r{&################\tabby&\hfil################\unskip}
    \def\c{&################\tabby&\hfil################\unskip\hfil}
    \def\l{&################\tabby&################\unskip\hfil}
    \edef\templ{\noexpand\vdimens ########\unskip  #1 
         \unskip&########\tabskip=0pt&########\cr}
    \tabledelim
    \edef\body##1{ \vbox{
        \tabskip=0pt \offinterlineskip
        \halign to \twidth {\templ ##1}}} }

\newbox\grsign \setbox\grsign=\hbox{$>$}
\newdimen\grdimen \grdimen=\ht\grsign
\newbox\laxbox \newbox\gaxbox
\setbox\gaxbox=\hbox{\raise.5ex\hbox{$>$}\llap
	{\lower.5ex\hbox{$\sim$}}}\ht1=\grdimen\dp1=0pt
\setbox\laxbox=\hbox{\raise.5ex\hbox{$<$}\llap
	{\lower.5ex\hbox{$\sim$}}}\ht2=\grdimen\dp2=0pt
\def\simlt{\mathrel{\copy\laxbox}}
\def\simgt{\mathrel{\copy\gaxbox}}

\def\uJy{\ifmmode{\,\mu{\rm Jy}}\else$\,{\mu{\rm Jy}}$\fi}
\def\mJy{\ifmmode{\,{\rm mJy}}\else${\,{\rm mJy}}$\fi}
\def\MHz{\ifmmode{\,{\rm MHz}}\else{$\,{\rm MHz}$}\fi}
\def\GHz{\ifmmode{\,{\rm GHz}}\else{$\,{\rm GHz}$}\fi}
\def\solar{\ifmmode_{\mathord\odot}\else$_{\mathord\odot}$\fi}
\def\Msolar{\ifmmode{\, {\rm M\solar}}\else{${\, {\rm M\solar}}$}\fi}
\def\Rsolar{\ifmmode{\, {\rm R\solar}}\else{${\, {\rm R\solar}}$}\fi}
\def\kms{\ifmmode{\,{\rm km\,s^{-1}}}\else${\,{\rm km\,s^{-1}}}$\fi}
\def\kpc{\ifmmode{\,{\rm kpc}}\else${\,{\rm kpc}}$\fi}
\def\us{\ifmmode{\,\mu{\rm s}}\else$\,{\mu{\rm s}}$\fi}
\def\ms{\ifmmode{\,{\rm ms}}\else$\,{{\rm ms}}$\fi}
\def\y{\ifmmode{\,{\rm y}}\else$\,{\rm y}$\fi}
\def\h{\ifmmode{^{\rm h}}\else$^{\rm h}$\fi}
\def\m{\ifmmode{^{\rm m}}\else$^{\rm m}$\fi}
\def\s{\ifmmode{^{\rm s}}\else$^{\rm s}$\fi}
\def\Lmin{\ifmmode{L_{min}}\else{$L_{min}$}\fi}

\input psfig.sty


\overfullrule=0pt


\singlespace



\font\lgh=cmbx10 scaled \magstep2
\def\grb{GRB\thinspace 980425}
\def\1SAXJ1935{1SAX~J1935.3$-$5252}
\def\J1935{J193503.3$-$525045}
\def\hb{\hfill\break}

\def\localization{1}
\def\lightcurves{2}
\def\spectralindex{3}
\def\ISS{4}
\def\TB{5}

\def\fluxtable{1}


\hrule
\bigskip
\line{\lgh The $\gamma$-ray burst of 980425 and its association \hb}
\line{\lgh with the extraordinary radio emission from \hb}
\line{\lgh a most unusual supernova\hb}

\bigskip

\line{S. R. Kulkarni$^1$, D. A. Frail$^2$, M. H. Wieringa$^3$,
R. D. Ekers$^4$, E. M. Sadler$^5$,\hb} 
\line{ R. M. Wark$^3$, J. L. Higdon$^3$, E. S. Phinney$^6$, and
J. S. Bloom$^1$\hb}

\bigskip 

\line{$^1$ Division of Physics, Mathematics \&\ Astronomy,
Caltech 105-24, Pasadena CA 91125, USA\hb} 

\line{$^2$ National Radio Astronomy Observatory, Socorro, NM, 87801,
USA\hb}

\line{$^3$ Paul Wild Observatory, Locked Bag 194, Narrabri NSW 2390,
Australia\hb}

\line{$^4$ Australia Telescope National Facility, CSIRO,
P.O. Box 76, Epping, NSW 2121, Australia\hb}

\line{$^5$ School of Physics, University of Sydney, NSW 2006,
Australia\hb}

\line{$^6$ Division of Physics, Mathematics \&\ Astronomy,
Caltech 130-33, Pasadena, CA 91125, USA \hb}
\medskip
\hrule
\bigskip

\noindent

{\it This manuscript has been been submitted to Nature on 29 June 1998.
We are making this available via astroph given the intense interest
in SN 1998bw and its relation to GRB 980425. You are free to refer to
this paper in your own paper. However, we do place restrictions
on any dissemination in the popular media.  The article is under embargo
until it is published. For further enquiries please contact Dale Frail
(dfrail@nrao.edu) or Shri Kulkarni (srk@astro.caltech.edu). }

\vfill\eject

\vfill\eject

{\bf Supernova SN 1998bw exploded in the same direction and at 
about
the same time as the gamma-ray burst GRB 980425.  Here we report radio
observations of this type Ic supernova, beginning 4 days
after the gamma-ray burst.  At its peak the radio source is the most
luminous ever seen from a supernova, $\nu L_{\nu}=4\times
10^{38}\,\hbox{erg s}^{-1}$ at 5 GHz.  More remarkably, the
traditional synchrotron interpretation of the radio emission requires
that the radio source be expanding at an apparent velocity of at least
twice the speed of light, indicating that this supernova was
accompanied by a shock wave moving at relativistic speed.  The energy
$U_e$ associated with the radio-emitting relativistic electrons must
lie between $10^{49}\,\hbox{erg}<U_e<10^{52}\,\hbox{erg}$, and thus
represents a significant fraction of the total kinetic energy $\sim
10^{51}\,\hbox{erg}$ associated with supernova explosions.  The
presence of a relativistic shock in SN 1998bw suggests a physical
connection with the gamma-ray burst GRB 980425.  We argue that this
represents a second class of gamma-ray burst, with much lower emitted
energy $\sim 10^{48}\,\rm erg$ in gamma-rays than the two powerful
$\sim 10^{53} \,\rm erg$ high-redshift gamma-ray bursts previously
identified.}

It is only within the last year that we have begun to understand GRBs.
At least one GRB is known to be securely of extragalactic
origin\refto{met97} and the energy release, $E_0$, of another GRB has
been estimated to be as high as $10^{53}$ erg (ref. \Ref{srk98}).  This
revolution in our understanding is due to
accurate localization of GRBs by the Italian-Dutch satellite
BeppoSAX\refto{Boella97} and the discovery of relatively long-lived
emission at X-ray\refto{cos97}, optical\refto{van97} and radio
wavelengths\refto{Frail97} -- the so-called ``afterglow'' phenomenon.
The burst itself and especially the afterglow emission are nicely
accounted for by  ``fireball''
models\refto{Meszaros97,Vietri97,Waxman97} which are similar to
supernova models but with material (``ejecta'') moving at relativistic
speeds. The inferred mass of the ejecta ($M_{ej}$) is an astonishingly
small $10^{-5}\,M_\odot$.  Speculations for the cause of this
explosion abound but most of them involve the formation of a black
hole. In one set of models, the black hole is the end product of
coalescence of neutron stars and in another model the black hole is the
end product of a massive star.  The reader is referred to refs.
\Ref{Fishman95,Piran97} for overviews.

\refis{met97}
	Metzger, M. R. {\it et al.} 
	Spectral constraints on the redshift of the optical 
	counterpart to the gamma-ray burst of 8 May 1997. 
	{\it Nature} {\bf 387}, 878-880 (1997).

\refis{srk98}
	Kulkarni, S. R. {\it et al.} 
	Identification of a host galaxy at
	redshift z = 3.42 for the $\gamma$-ray burst of December 1997. 
	{\it Nature} {\bf 393}, 35-39 (1998).

\refis{Boella97}
	Boella G.  {\it et al.},
	BeppoSAX, the wide band mission for x-ray astronomy.
	{\it Astron. Astrophys. Suppl. Ser.} {\bf 122}, 299-399 (1997).

\refis{cos97} 
        Costa, E. {\it et al.} 
	Discovery of an X-ray afterglow associated with the gamma-ray 
	burst of 28 February 1997. 
	{\it Nature} {\bf 387}, 783-785 (1997).

\refis{van97}
        Van Paradijs, J. {\it et al.} 
	Transient optical emission from the error box of the 
	$\gamma$-ray burst of 28 February 1997. 
	{\it Nature} {\bf 368}, 686-688 (1997).

\refis{Frail97} 
        Frail, D. A., Kulkarni, S. R., Nicastro, L., Feroci, M. \&\ 
        Taylor, G.  B.  
        The radio afterglow from the gamma-ray burst of 8 May 1997.
        {\it Nature} {\bf 389}, 261-263 (1997).

\refis{Meszaros97}   
	M\'esz\'aros, P. \&\ Rees, M. J.
	Optical and long-wavelength afterglow from gamma-ray
	bursts.
	{\it Astrophys. J.} {\bf 476}, 232-237 (1997).
\refis{Vietri97} 
	Vietri, M.
	The afterglow of gamma-ray bursts: the cases of GRB
	970228 and GRB 970508.
	{\it Astrophys. J.} {\bf 488}, L105-L108 (1997).

\refis{Waxman97}    
	Waxman, E.
	Gamma-ray-burst afterglow: supporting the cosmological
	fireball model, constraining parameters, and making
	prediction.
	{\it Astrophys. J.} {\bf 485}, L5-L8 (1997).

\refis{Fishman95}
	Fishman, G. J. \&\ Meegan, C. A. 
	Gamma-ray bursts.
	{\it Annu. Rev. Astron. Astrophys.} {\bf 33}, 415-458 (1995).

\refis{Piran97} Piran, T.
	Towards understanding gamma-ray bursts.
	in {\it Unsolved problems in astrophysics}, Eds.
	J. N. Bahcall \&\ J. P. Ostriker, Princeton University Press,
	343-377 (1997).


Like GRBs, supernovae (SNe) are also explosive events, but are much
more numerous.  The observed light curves and spectra are used to
classify supernovae and while it is not firmly established,
astronomers believe that SNe of Type II, Ib and Ic mark the death of a
massive star resulting in the formation of a neutron star or a black
hole. SNe of type Ia are popularly attributed to the destruction of a
massive white dwarf.  The bulk of the observable energy is initially
in the form of the kinetic energy of the ejected mass $M_{ej}$,
$1M_\odot\simlt M_{ej}\simlt 20M_\odot$.  Estimates of the total
kinetic energy $E_0$ for SNe appear to cluster in a narrow range
around $10^{51}$ erg. The ejecta move at speeds $\sim
\sqrt{2E_0/M_{ej}} \sim 10^9$ cm s$^{-1}$ 
and the resulting
shocks produce X-ray and optical emission.

Three parameters are sufficient to model the evolution of an
astronomical explosion, be it SN or GRB: the kinetic energy released
($E_0$), the mass (``ejecta'') in which this energy is initially
deposited, $M_{ej}$, and the density of the ambient gas.  The primary
difference between SNe and GRBs lies in the the mass and hence in the
speed and optical depth of the ejecta: high optical depth and
non-relativistic speeds $\beta c$ in SNe, $\beta\simlt 0.2$ or a bulk
Lorentz factor, $\Gamma\simlt 1.02$; and low optical depth and
relativistic speeds in GRBs with bulk Lorentz factor, $\Gamma\sim
300$. Here, following the standard convention, $\Gamma  \equiv (1 -
\beta^2)^{-1/2}$. This difference accounts for the efficient high
energy emission of GRBs and the lower efficiency and lower energy
(optical) emission of SNe. Secondarily, while the total energy
released is about the same in SNe and GRBs ($E_T\sim 10^{53.5}$ erg),
in normal SNe, 99\%\ of that energy is carried off by neutrinos. And
the high optical depth means that 99\% of the remaining 1\% is
converted into kinetic energy during adiabatic expansion. At
interstellar particle densities $< 10\,\hbox{cm}^{-3}$, only $10^{-4}$
of $E_T$ escapes in electromagnetic emission during the early months,
the rest of the kinetic energy being radiated by the supernova remnant
over the following $\sim 10^6 \,\rm y$.

In this paper we present radio observations of SN 1998bw (ref.
\Ref{Galama98}), the analysis of which leads us to suggest that this
SN is a link between some GRBs and supernovae.  SN 1998bw appears to
have exploded within the joint time and spatial error box of GRB
980425 (ref. \Ref{Soffitta98}).  The low probability ($10^{-4}$) of
finding such a young supernova within the compact error
circle\refto{Soffitta98} of the GRB suggests an
association\refto{Galama98} between SN 1998bw and GRB 980425.  As
discussed in detail below, the radio emission from SN 1998bw has
several unusual properties, including being the most
luminous\refto{Wieringa98} radio supernova ever observed.  These
unique features further strengthen\refto{Sadler98} the case for
association.  Inspired by this identification, other coincidences
between Type Ib/Ic SNe and gamma--ray bursts have been
noted\refto{Wang98,Woosley98}.  Thus SN 1998bw may not be the only
example of this potentially new class of GRBs.

\refis{Galama98}
        Galama, T. J. et al.
        Discovery of the peculiar supernova 1998bw in the error box
        of GRB 980425.
        astro-ph/9806175, http://xxx.lanl.gov (1998).

\refis{Soffitta98} 
        Soffitta, P. et al.
        {\it Intl. Astron. Circ.} {\bf 6884}, (1998).

\refis{Wieringa98}
	Wieringa, M., Frail, D. A., Kulkarni, S. R., Higdon, J. L.,
	Wark, R. \&\ Bloom, J. S. 
        {\it Intl. Astron. Circ.} {\bf 6896}, (1998).
	
\refis{Sadler98}
	Sadler, E. M., Stathakis, R. A., Boyle, B. J. \&\
	Ekers, R. D.
        {\it Intl. Astron. Circ.} {\bf 6901}, (1998).

\refis{Wang98}
	Wang, L.  \&\ Wheeler, J. C. 
	The supernova-gamma-ray burst connection.
	http://xxx.lanl.gov, astro-ph/9806212 (1998).

\refis{Woosley98}
	Woosley, S. E., Eastman, R. G. \&\ Schmidt, B. P.
	Gamma-ray bursts and Type Ic supernovae: SN 1998bw.
	http://xxx.lanl.gov, astro-ph/9806299 (1998).

The arguments listed above in favor of the GRB--SN association are
compelling but fail to provide an observational clue as to how a SNe
can generate a burst of gamma-rays. Theoretical
models\refto{Woosley98} have great difficulty generating a burst of
gamma-rays.  In this paper we show how a simple interpretation of the
radio data forces us to conclude that SN 1998bw had a shock moving at
relativistic speed, ahead of the low velocity ejecta which powers the
optical light curve. If our interpretation is correct (and fortunately
it is amenable to observational verification) then we have identified
a specific phenomenon -- a relativistic shock -- which could
potentially generate a burst of gamma-rays at early times in a SN.

\noindent{\bf GRB 980425}  

Soffitta et al.\refto{Soffitta98} detected a gamma-ray burst of
$\sim$30 seconds duration in the BeppoSAX Gamma-Ray Burst Monitor and
the Wide Field Camera (WFC) on 1998 April 25.90915 UT. Follow-up
observations of the 8-arcmin (radius) WFC error circle were made
beginning on April 26.31 UT (ref. \Ref{Pian98g}) and May 2.60 UT
(ref. \Ref{Pian98b}) with the BeppoSAX Narrow Field Instruments
(NFI). Two previously unknown X-ray sources were found, one of which
(1SAXJ1935.0$-$5248) remained steady between the two epochs. The
other, 1SAXJ1935.3$-$5252, was detected at (2.4$\pm0.5)\times 10^{-3}$
count~s$^{-1}$ (1.6-10 keV) during the first 27.7 hrs of the April 26
observation, but was not detected during the next 16.6 hrs. The
3-$\sigma$ upper limit was 1.8$\times{10}^{-3}$ count~s$^{-1}$.  Nor
was it detected during the May 2 observation and the corresponding
upper limit is 1.5$\times{10}^{-3}$ count~s$^{-1}$.

\refis{Pian98g}
        Pian, E. et al.
        {\it GCN Note} No. 61 (1998a).

\refis{Pian98b}
        Pian, E., Frontera, F., Antonelli, L. A. \&\ Piro, L.
        {\it GCN Note} No. 69 (1998b).

Thus initial searches for optical and radio afterglow from the
gamma-ray burst concentrated on the NFI position of this variable
source 1SAXJ1935.3$-$5252 (see Figure~\localization).  Galama et
al.\refto{Galama98a} and Bloom et al.\refto{Bloom98} reported the
absence of any transient optical source brighter than R$<$21 mag with
changes larger than $\pm$0.2 mag.

\refis{Galama98a} 
        Galama, T. J. et al. 
        {\it GCN Note} No. 62, (1998).

\refis{Bloom98}
        Bloom, J. S., Kulkarni, S. R., Djorgovski, S. G.,
        McCarthy, P. \&\ Frail, D.
        {\it GCN Note} No. 64, (1998).

We initiated our program of radio observations at the Compact Array
(ATCA), an interferometric East-West array operated by the Australia
Telescope National Facility. Observations began on April 28.73 in the
6-cm and the 3-cm bands.  From the analysis of data of April 28 and
April 29 we reported\refto{Wieringa98} the absence of any radio source
in the NFI error circle of 1SAXJ1935.3$-$5252
with a 3-$\sigma$ upper limit of 0.3
mJy.  Similarly, by averaging the datasets from the May 10 to June 22
monitoring effort, we can derive upper limits of 0.4, 0.2 and 0.4 mJy
in the 20-cm, 6-cm, and 3-cm bands, respectively.
This sensitivity would have been
sufficient to have detected the radio emission seen from GRB 970508
(ref. \Ref{Frail97}) and the two recently detected radio afterglows,
GRB 980329 (ref.  \Ref{Taylor98}) and GRB 980519 (ref. \Ref{Frail98}).
However, to date, we have detected only three radio afterglows out 
of twelve well localized-bursts. Thus the failure to find 
a radio afterglow of this GRB is not alarming or constraining.

\refis{Wieringa98}
        Wieringa, M., Frail, D. A., Kulkarni, S. R.,
        Higdon, J. L. \&\ Wark, R.
        {\it GCN Note} No. 63, (1998).

\refis{Taylor98}
        Taylor, G. B, Frail, D. A., Kulkarni, S. R., \& Shepherd, D. S.
        Discovery of the radio afterglow from the optically dim
        gamma-ray burst of March 29, 1998.
        {\it Astrophys. J.}, in press, (1998).

\refis{Frail98}
        Frail, D. A., Taylor, G. B. \&\ Kulkarni, S. R.
        {\it GCN} Note No. 89, (1998).

\noindent{\bf SN 1998bw}  

Inside the larger 8-arcmin field-of-view of the WFC, but not
coincident with either of the two NFI X-ray sources, Galama et
al.\refto{Galama98b} noted a R=15.7 mag object located on the western
spiral arm of the barred spiral galaxy ESO 184-G82. This bright object
was not present in the Digitized Sky Survey, and, furthermore
continued to brighten. Consequently, Galama et al.\ suggested that the
source was a possible supernova (SN).  Spectra taken by Lidman et
al.\refto{Lidman98} appeared to rule out a type-II or type-Ia
supernova.  Sadler et al.\refto{Sadler98} pointed out its similarities
to the pre-maximum spectrum of SN 1998bw with the type-Ib supernova
1983N, while Patat and Piemonte\refto{Patat98} noted that its helium
lines were weak or absent, making it more like the spectra seen in
type-Ic supernova. Given these observational developments, the Central
Bureau for Astronomical Telegrams designated this object SN 1998bw. In
Figure \localization, we summarize the various localizations.

\refis{Galama98b}
        Galama, T. J., Vreeswijk, P. M., Pian, E., Frontera, F.
        Doublier, V. \&\ Gonzalez, J. -F.
        {\it Intl. Astron. Union} {\bf 6895}, (1998).

\refis{Lidman98}
        Lidman, C. et al.
        {\it Intl. Astron. Union} {\bf 6895}, (1998).

\refis{Sadler98}
        Sadler, E. M., Stathakis, R. A., Boyle, B. J.,
        \&\ Ekers, R. D.
        {\it Intl. Astron. Union} {\bf 6901}, (1998).

\refis{Patat98}
        Patat, F. \&\ Piemonte, A.
        {\it Intl. Astron. Union} {\bf 6918}, (1998).

Galama et al.\refto{Galama98} present an optical (UBVRI) light curve
for SN~1998bw.  They find a peak absolute magnitude in the B band of
$-18.9$ magnitude and remark that it is unusually luminous for a type
Ib/Ic supernova\refto{mb90}. However, there is considerable scatter in
the peak luminosity of Type Ib/Ic. Indeed, the type Ic SN 1992ar was a
magnitude brighter\refto{Hamuy92} than SN 1998bw.  On the other hand,
as noted above, the optical spectrum of SN 1998bw is certainly
unusual.  Further unusual properties are the late-time spectrum, and
indications of abnormally high-velocity gas.  At twenty days past
maximum light the optical spectrum has weaker-than-normal and
broader-than-normal absorption lines\refto{Stathakis98}.  R. A.
Stathakis (pers. comm.) notes that ejection velocities measured from
the blue wings of the Ca~II line are as high as 60,000 km s$^{-1}$ at
the beginning of May and 30,000 km s$^{-1}$ in mid--May, 1998.

\refis{Stathakis98}
	Stathakis, R. A. et al. 
	Spectroscopy of SN 1998bw.
	in prep., (1998).

From optical spectroscopic observations\refto{Lidman98,Tinney98} the
redshift to ESO 184-G82 is measured to be 0.0083 (heliocentric).  The
detection of narrow line absorption from sodium D
lines\refto{Stathakis98} at the redshift of ESO 184-G82 shows that the
SN is either in or behind ESO 184-G82.  The distance to this galaxy,
assuming a Hubble constant of 65 km s$^{-1}$ Mpc$^{-1}$, is 38 Mpc.
The SN 1998bw, even at an assumed distance of 38 Mpc, is already ten
times brighter than a typical Type Ib/Ic SN. Thus we argue that SN
1998bw is unlikely to be an unrelated background SN (ie.~even more
distant) but indeed is located within ESO 184-G82.

\refis{Tinney98} 
        Tinney, C., Stathakis, R., Cannon, R. \&\ Galama, T.
        {\it Intl. Astron. Union} {\bf 6896}, (1998).

\refis{mb90} 
Miller, D. L., \& Branch, D. 
Supernova absolute-magnitude distributions. 
{\it Astron. J.} {\bf 100}, 530-539 (1990).

\refis{Hamuy92}
	Hamuy, M. et al.
	{\it Intl. Astron. Union Circ.} {\bf 5574}, (1992).

\noindent{\bf Radio Observations of SN 1998bw}

In the course of searching for radio emission from \1SAXJ1935\ we
noted\refto{Wieringa98b} that the brightest radio source in the WFC
error circle, \J1935\ coincided with the supernova candidate of Galama
et al. and that it too brightened considerably by May 5.6 UT. This
strong early detection of a SNe motivated us to begin a radio
monitoring program initially in the 6-cm (centre frequency, 4800
MHz) and the 3-cm (8640 MHz) bands 
at the position of the X-ray transient.  From
May 7 we also observed in the 20-cm (1384 MHz) and the 13-cm (2496
MHz) bands and shifted the field center $3'$ NW to the position of SN
1998bw.  The results of this monitoring effort are summarized in Table
\fluxtable\ and Figures
\lightcurves\ and \spectralindex.

\refis{Wieringa98b}
Wieringa, M. {\it et al.} {\it Intl. Astron. Circ.} {\bf 6896}, (1998).

The flux in the 3-cm band and in the 6-cm band rise approximately
linearly with time.  The rapid rise enables us to empirically
establish the epoch of the origin of the radio emission to better
precision than that obtained from optical
measurements\refto{Galama98}.  As can be seen from Figure
\lightcurves, this epoch is close ($\pm$2 days) to the time of the
gamma-ray burst.  This further strengthens the case for a physical
association between the GRB and the SN.  Given the near coincidence of
the initiation of the radio emission and the time of the GRB we assume
that the epoch of radio emission was the same as that of GRB 980425.

A long ATCA observation (9 hrs) was obtained on day 10.  The 6-cm flux
showed a smooth increase from 37 mJy to 41.6 mJy, consistent with the
overall rising flux.  At the same time the 3-cm flux increased from 46
mJy to 57 mJy but with considerable (20\%) scatter. The scatter could
be due to pointing errors, since SN 1998bw was located on the
half-power response of the 3--cm primary beam for this observation.  On
day 12 we obtained 11 observations spread over 7 hrs with SN 1998bw at
the field centre. The variations in the flux are no larger than 6\% at
6 cm and 3\% at 3--cm.  The almost daily observations from day 12 to 22
with SN 1998bw at the field centre show a smooth change in flux density
with day-to-day variations $<5$\% (20-cm) and $<1$\% (13-cm).

A single observation was also made on the night of 1998 May 7.7 UT,
using the SCUBA bolometer array on the 15-m James Clark Maxwell
Telescope (JCMT). Although SN 1998bw was observed at low elevation
($<17^\circ$) it was detected in the 2-mm band with a flux density of
39$\pm$11 mJy. At 1.35 mm the value is $-$21$\pm$32 mJy. We note that
the 2-mm data point was taken very close to the peak of the 6-cm and
3-cm emission.

The flux reaches a maximum on day 12 with flux densities,  S$(\nu)$ of
49 mJy (in the 3-cm band) and 45 mJy (6-cm band).  The emission decays
on a timescale similar to that of the rise-timescale.  This rapid flux
evolution is mirrored in the spectral index ($\alpha$) plot (Figure
\spectralindex) with the largest changes occurring in the first 20 days;
here $\alpha$ is the power law spectral index defined by the equation
$S(\nu) \propto\nu^\alpha$. When the flux in the 6-cm band peaks, the
spectral index from 20-cm to 6-cm is approximately 2.

After day 20, a new component of emission emerged. A broad maximum was
reached at 3 cm (23 mJy) and 6 cm (30 mJy) on day 30.  The 20-cm light
curve rose $\sim$0.5 mJy day$^{-1}$ from day 12, reached a peak (29
mJy) on day 45 and thereafter decayed. Beyond day 35 the 3-cm, 6-cm
and 13-cm fluxes have begun a simple decay, $\sim$0.5 mJy day$^{-1}$.
During this time the spectral index, as measured between pairs of
adjacent wavelengths, is asymptotically converging to a value between
$-0.5$ to $-1.0$ (Figure \spectralindex).


\noindent{\bf Comparison of RSN 1998bw with other radio supernovae} 

Superficially the radio light curve in Figure \lightcurves\ resembles
other radio SNe (hereafter we will use the abbreviation RSNe for radio
SNe).  However, there are significant differences.  In order to make a
proper comparison of this RSN with other RSNe, we now summarize the
known properties of RSNe.  The reader is referred to Weiler et
al.\refto{Weiler86} and Weiler \&\ Sramek\refto{Weiler88} for recent
observational reviews and Chevalier\refto{Chevalier82} for a
theoretical overview.

\refis{Chevalier82}
        Chevalier, R. A.
        The radio and X-ray emission from type II supernovae.
        {\it Astrophys. J.} {\bf 259}, 302-310 (1982).

\refis{Weiler86} 
        Weiler, K. W., Sramek, R. A., Panagia, N., van der Hulst, J. M.,
        \&\ Salvati, M.
        Radio Supernovae.
        {\it Astrophys. J.} {\bf 301}, 790-812 (1986).

\refis{Weiler88} 
        Weiler, K. W. \&\ Sramek, R. A.
        Supernovae and supernova-remnants.
        {\it Rev. Astron. Astrophys.} {\bf 26}, 295-341 (1988).

Radio emission is seen only from Type II or Type Ib/Ic SNe.  While only
five type Ib and Ic SNe have been studied in any detail, they all
exhibit a more rapid radio flux density evolution than type-II SNe. At
centimeter wavelengths the Ib/Ic RSNe peak some 10 to 40 days after the
initial explosion, while type II RSNe peak from 50 to 1000 days.  Thus
from radio observations alone we can ascertain that SN 1998bw is not a
type II RSN but a type Ib/Ic RSN.

The 6-cm flux ($S_6$) peaks on day 12 and the spectral luminosity,
$4\pi S_6d^2$ is $\sim 8\times 10^{28}d_{38}^{2}$ ergs s$^{-1}$ Hz$^{-1}$;
here and below we assume that SN 1998bw is located in the galaxy
ESO~184-G82 at a distance of $38d_{38}$ Mpc. According to Weiler et
al.\refto{Weiler98} RSNe of type Ib and Ic are standard candles with
peak 6-cm spectral luminosity around 
$1.9\times 10^{27} \hbox{erg s}^{-1}\hbox{Hz}^{-1}$
(Type Ib) and $6.5\times 10^{26} \hbox{erg s}^{-1}\hbox{Hz}^{-1}$
(Type Ic). We caution the reader that this conclusion
is based on only five Type Ib/Ic radio SNe, and the data for some of
these is quite sparse.  Nonetheless,  RSN 1998bw is more luminous than any
previously studied RSNe including the powerful type II SN 1988Z
(ref. \Ref{vandyk93}; redshift, $z=0.022$) which achieved a peak 6-cm
spectral luminosity of $2\times 10^{28} \hbox{erg s}^{-1}\hbox{Hz}^{-1}$
three years after the initial explosion.

\refis{Weiler98}
        Weiler, K. W., van Dyk, S. D., Montes, M. J.,
        Panagia, N., Sramek, R. A.
        Radio supernovae as distance indicators.
        {\it Astrophys. J.} in press,  (1998).

\refis{vandyk93}
        van Dyk, S. D., Weiler, K. W., Sramek, R. A. \&\
        Panagai, N.
        SN 1988Z: the most distant radio supernova.
        {\it Astrophys. J.} {\bf 419}, L69-L72 (1993).

On longer timescales, RSN 1998bw continues to follow a pattern
different from that of other RSNe.  As stated earlier around about day
20, a new component appears.  This component peaks at 6-cm at
about day 30 after which the peak moves to longer frequencies whilst
maintaining approximately the same flux density. The component peaks at
20-cm on about day 45.  This frequency-dependent evolution ensures that
the spectral index is not constant across the cm-wave band as can be
seen in Figure \spectralindex.  In contrast, at late times, the radio
emission from previously known RSNe decays monotonically at all
frequencies.  This broad-band decay is interpreted in terms of an
expanding optically thin synchrotron emission source; the expansion
leads to the overall decay and the optical thin condition ensures that
the spectral index is constant during the expansion.  By day 50, it is
clear that the flux in the shortest two wave-bands (3-cm and 6-cm) are
declining. The spectral index between these two bands is $\sim -1$
and can be reasonably interpreted as the spectral index of the
optically thin portion of the synchrotron spectrum. Our measured
spectral index is comparable to those measured in type Ib/Ic RSNe but
different from the $-0.5$ to $-0.7$ of Type II
RSNe\refto{Weiler86,vandyk93}.   Again, this result emphasizes that RSN
1998bw is not a type II RSNe.

\noindent{\bf Constraint on the size from variability} 

The expected angular size of the expanding radio photosphere
$\theta_S$ can be estimated by assuming that the radio photosphere
expands at the same speed as that inferred from optical spectroscopy
which we noted earlier\refto{Stathakis98} to be $v_O\sim 6\times 10^9$
cm. We obtain $\theta_S=0.91t_d v_{60} d_{38}^{-1}\,\mu$arcsec where
$v_{60}$ is the assumed velocity of expansion in units of 60,000 km
s$^{-1}$ and $t_d$ is time since explosion in units of days. Thus the
expected angular radius on day 10 is about $9\,\mu$arcsec. This is
sufficiently small that one expects to see deep modulation of the
received signal on timescales of hours, due to
scattering\refto{narayan92,good97,Walker98} of the radio waves by
density irregularities in the diffuse ionized medium of our Galaxy.

\refis{narayan92}
	Narayan, R.
	The physics of pulsar scintillation.
	{\it Phil. Trans. R. Soc. Lond.} A {\bf 341}, 151-165 (1992).

\refis{good97}
	Goodman, J. 
	Radio scintillation of gamma-ray-burst afterglows. 
	{\it New Astro.}, {\bf 2}, 449-460 (1997).

\refis{Walker98}
	Walker, M. A. 
	Interstellar scintillation of compact extra-galactic
	radio sources. {\it M.N.R.A.S.} {\bf 294}, 307-311 (1998).

For the purpose of theoretical modelling, 
the Galactic ionized medium can be conveniently approximated by a
thin screen located at an effective distance of $D\sim 0.5$--1 kpc.
Scattering is considered to be in the ``weak'' regime when the Fresnel
scale, $r_F=\sqrt{\lambda D/2\pi}$ is smaller than $r_0$, the spatial
scale of the irregularities in the scattering screen; here $\lambda$ is
the wavelength of observations. In this regime, the intensity
variations are modest and $m$, the modulation index (defined as the
ratio of the rms of the intensity variations to the mean) is well below
unity. When $r_0<<r_F$, the scattering is considered to be in the
``strong'' regime and the multitude of $r_0$-sized patches within a
single Fresnel scale result in multi-path propagation. The result is
deep intensity variations.

Both $r_0$ and $r_F$ are functions of the observing frequency. Instead
of talking in terms of $r_0$ and $r_F$ it is more convenient to talk
in terms of the transition frequency\refto{Walker98}, $\nu_0$.
Observations at frequency ($\nu$) above $\nu_0$ are in the weak regime
and observations at lower frequency suffer from strong scattering.
Assuming typical interstellar medium parameters for this
direction\refto{Cordes93}, $\nu_0$ is in the frequency range 3--7 GHz,
a frequency range nicely straddled by our ATCA observations.  The
Fresnel angle, $\theta_F\equiv r_F/D$ evaluated at the transition
frequency of (say) 5 GHz is $\theta_0\sim 4$ $\mu$arcsec; here and
below we have set $D$ to 1 kpc.

\refis{Cordes93}
	Taylor, J.H. \&\ Cordes, J.M.
	Pulsar distances and the galactic distribution of free electrons.
	{\it Astrophys. J.} {\bf 411}, 674-684 (1993).

Interstellar scattering allows us to constrain source size as long as
the source smaller than one of three
characteristic angular scales:  (i) ``weak'',
$\theta_W=\theta_0(\nu_0/\nu)^{1/2}$, (ii) ``refractive'',
$\theta_R=\theta_0(\nu_0/\nu)^{11/5}$ and (iii) ``diffractive'',
$\theta_D=\theta_0(\nu_0/\nu)^{-6/5}$.  
In Figure \ISS\ we display the various regimes of interstellar
scattering. We immediately note that diffractive scattering may have
been important only at the earliest times. Unfortunately, during this
time SN 1998bw was not located at the field center and antenna
pointing errors likely dominate the observed variations. Outside these
times we are in the weak scattering or refractive scattering regimes.

The expected modulation index and the timescale for the variability is
given in refs. \Ref{good97,Walker98}.  On day 12 the 3-cm and 6 cm
data showed less than 3\% and 6\%, respectively, over 7 hrs. Using an
upper bound of three times the observed fractional variation we find
that a lower bound on the velocity of expansion of the radio
photosphere is 70,000-90,000 km s$^{-1}$. However, better constraints
come from the lack of variability at the low frequencies. This is best
seen in Figure \ISS\ in which we see the 20-cm and 13-cm lie squarely
in the refractive regime. Beginning at day 12 the fluxes at 20-cm and
13-cm rise smoothly over the next 10 days. After 
subtracting the  smooth rise  we find the 
day-to-day variations of 5\% at 20-cm
and 1\% at 13 cm 
(over a period of 7 days). 
The expected modulation index depends on the assumed
$\nu_\circ$, however, as can be seen in Figure \ISS\ the 13 cm and 20
cm datapoints nicely bracket the assumed uncertainty in
$\nu_\circ$. The modulation index over this range is expected to be
$\sim$35\% (ref. \Ref{good97}). From the theory of refractive
scintillation we infer the expansion of the radio photosphere to be
$>>v_o$, and place a lower limit of 0.3$c$.

As a cross-check on these estimates we note that the radio afterglow of
GRB\thinspace{970508} which is located at a similar Galactic latitude
exhibited strong variability at 8.46 and 4.86 GHz for at least three
weeks, from which it was possible to infer its angular size and deduce
the expansion of this cosmological GRB fireball (Frail et al. 1997).
Thus the absence of any scintillation in \J1935\ is due to the rapid angular
expansion of this relatively nearby radio source.

\noindent{\bf Modelling the radio light curves}

Radio observations of supernovae have been interpreted in the
framework of the ``mini-shell'' model\refto{Chevalier82}.  The ejecta
push a decelerating shock ahead of themselves into the circumstellar
material.  The shocked circumstellar material is then subject to
Rayleigh-Taylor instability.  This instability can drive turbulent
motions which amplify magnetic fields and help
Fermi-accelerate\refto{Ball95} particles to relativistic energies.
Accelerated electrons gyrate in the newly amplified magnetic field and
generate strong synchrotron emission.

\refis{Ball95}
	Ball, L. \&\ Kirk, J.G.
	The acceleration of electrons in the radio supernova SN 1986J.
	{\it Astron. Astrophys.} {\bf 303}, L57-L60 (1995).

The fact that the 
observed radio flux density drops at low frequency
while the flux density of optically thin
synchrotron radiation would continue to increase with decreasing
frequency requires a frequency-dependent absorption mechanism. The two
dominant mechanisms are free-free absorption by the surrounding
circumstellar matter (assumed to be ionized by the SNe or the
progenitor) and synchrotron self-absorption.
Chevalier\refto{Chevalier82} concluded that the delayed turn-on of the
lower-frequency radio emission is best explained by free-free
absorption.  Subsequently, much of the RSNe data have been
modelled\refto{Weiler86,Weiler88} in the context of the minishell and
free-free absorption model.

However, Shklovskii\refto{Shklovskii85} noting the fast turn-on of SN
1983N, a type Ib RSN, suggested that synchrotron self-absorption was the
dominant absorption mechanism.  Recently, Chevalier\refto{Chevalier98}
has reconsidered this issue and concludes that both absorption
mechanisms may be important but that synchrotron self-absorption cannot
be ignored for type Ib/Ic RSNe.  Kulkarni \&\ Phinney\refto{Kulkarni98}
conclude that self-absorption is the dominant opacity in Type Ib/Ic
RSNe. However, it is also equally clear that free-free absorption is
manifestly important in Type II RSNe.

\refis{Kulkarni98}
        Kulkarni, S. R. \&\ Phinney, E. S.
        Relativistic shocks in radio supernovae: Robust 
        inference from brightness temperature.
        In preparation, (1998).

\refis{Shklovskii85}
        Shklovskii, I. S. 
        Synchrotron self-absorption of the radio emission of supernova
        1983.51.
        {\it Sov. Astron. Lett.} {\bf 11}, 105-106 (1985).

\refis{Chevalier98}
        Chevalier, R. A.
        Synchrotron self-absorption in radio supernovae.
        {\it Astrophys. J.} {\bf 499}, 810-819 (1998).

We now return to SN 1998bw. We noted earlier that the radio emission
is composed of two emission components (Figure \lightcurves).  The
first component, on about day 10, peaks in the 3-cm band with a flux
of 50 mJy.  This component has all the features of a classical
synchrotron self-absorbed spectrum.  Specifically on day 10 we infer
a peak frequency,
$\nu_p\sim 10$ GHz and associated peak flux $S_p\sim 50$ mJy.  It is well known
that in a homogeneous source with a powerlaw electron spectrum, the
self-absorbed synchrotron spectrum exhibits a $\nu^{5/2}$ power law
for frequency less than $\nu_p$. The observed spectral index of 2
between 20-cm and 3-cm of 2 (Table \fluxtable, Figure \spectralindex)
is in acceptable agreement with this expectation.  Having made the
case for a sychrotron self-absorbed spectrum we now apply diagnostics
of synchrotron theory to our data.

Kulkarni \&\ Phinney\refto{Kulkarni98} show that despite the
complication of external absorption, brightness
temperature ($T_B$) limits can be used to
draw robust conclusions from the light curves of RSNe ; here $T_B$ is
$$
 T_{B}(\nu) = \Bigl( {c^2\over 2 k}      \Bigr)
              \Bigl( {S(\nu)\over\pi \theta_S^2} \Bigr)
                \nu^{-2}. \eqno (1)
$$ 
The robustness is obtained from the well known
result\refto{Hoyle66,Kellerman68} that $T_B$ of a source radiating via
the incoherent synchrotron mechanism is limited to $T_{\rm icc}\simlt
10^{12}$ K.  The origin of this limiting $T_B$ is the
``inverse-Compton catastrophe''. When $T_B$ exceeds $T_{\rm icc}$
multiple inverse Compton scattering of the synchrotron radio photons
to $\gamma$-ray/X-ray energies begins to rapidly dominate the
luminosity.  Limits on the X-ray to $\gamma$-ray flux or plausible
total luminosities thus limit $T_B$ to $\simlt T_{\rm icc}$.

\refis{Hoyle66}
	Hoyle, F., Burbidge, G.R., \&\ Sargent, W.L.W. 
	On the nature of the quasi-stellar sources.
	Nature {\bf 209} 751-753 (1966).

\refis{Kellerman68}
        Kellerman, K. I. \&\ Pauliny-Toth, I. I. K.
        The spectra of opaque radio sources.
        {\it Astrophys. J.} {\bf 155}, L71-L78 (1968).

For our application, the precise value of $T_{\rm icc}$ does matter and
we now give an improved estimate of this. Assuming $\alpha=-1$ and $\nu_p=5$
GHz, following Readhead\refto{Readhead94}, we find  that ratio of the
inverse Compton ($L_{\rm ic}$) luminosity to the luminosity in
synchrotron photons ($L_{\rm synch}$) is given by
$$
	{L_{\rm ic}\over L_{\rm synch}} = \Bigl( {T_B\over 4\times
	10^{11} {\rm ~K}} \Bigr)^5 \Bigl[ 1+ \Bigl( {T_B\over 4\times
	10^{11} {\rm ~K}} \Bigr)^5 \Bigr].  \eqno (2)
$$
From the BeppoSAX NFI X-ray observations on day 1 and day 8 we infer an
upper limit (3$\sigma$) on the 1.6--10 keV flux from the SNe, $f_X
<10^{-13}$ erg cm$^{-2}$ s$^{-1}$.  On day 10, when the radio emission
peaks in the cm-wave radio band, the 6-cm flux, $S_R=50$ mJy and thus
the flux in the radio band is $S_R\nu_R\sim 2.5\times 10^{-15}$ erg
cm$^{-2}$ s$^{-1}$.  The ratio of the X-ray luminosity (if any) to the
radio luminosity is thus below 40 and using this value in Equation (2)
we obtain $T_{\rm icc} \simlt 5 \times 10^{11}$ K. A more precise value
depends on the cutoffs in the particle distribution and the $T_B$
adopted, which set the location of the ``Compton humps'' in the spectrum.

\refis{Readhead94}
        Readhead, A. C. S.
        Equipartition brightness temperature and the inverse
        Compton catastrophe.
        {\it Astrophys. J.} {\bf 426}, 51-59 (1994).

Assuming that the angular radius of the source $\theta_S=v_0t/d$ where
$v_0=$ 60,000$v_{60}$ km s$^{-1}$ is the assumed photospheric expansion
speed and using Equation (1) the inferred brightness temperature is
$T_B=2.0\times 10^{13} S({\rm mJy}) (\lambda/6\,{\rm cm})^2
v_{60}^{-2}t_d^{-2}d_{38}^2$ K; here $S$ is the flux at wavelength
$\lambda$. As can be seen in Figure \TB, if $v_0=60,000\,\hbox{km s}^{-1}$,
the inferred $T_B$ greatly exceeds $T_{\rm icc}$ at all the
observing frequencies.

Clearly, we have erred in one of our assumptions. What went wrong?  As
discussed earlier, the SNe cannot be in the foreground because of the
interstellar absorption at the red-shift of the host galaxy ESO 184-G82,
and if it is a background object the resulting $T_B$ as well as the
peak optical brightness would be even higher.  
The origin of the time for the radio emission is reasonably
secure (to within a day or so) as can be seen from looking at Figure
\lightcurves. Thus our assumed $v_0$ must be incorrect.

\noindent{\bf Relativistic Shocks}

The preceding argument forces us to
abandon the idea that the radio emitting region expands is coincident with
the optical photosphere. 
Let the expansion speed of the radio emitting front be $v=\beta c$; here
$v$ is the velocity of the expanding front and $c$ the speed of light.  Then
the apparent transverse expansion angular speed is $\Gamma\beta c$
where $\Gamma=(1-\beta^2)^{-1/2}$ is the Lorentz factor.
For a source at distance $d$ observed at time $t$ after the explosion, 
the specific intensity
$I_\nu=(S/\pi)(d^2/(\Gamma\beta c t)^2)=2kT_B\nu^2/c^2=2k{\cal D}T_B'(\nu')
({\cal D}\nu')^2/c^2$. The last equality follows because
$I_\nu \nu^{-3}$ is a Lorentz invariant, and the Doppler factor ${\cal D}
=[\Gamma(1-\beta\cos\theta)]^{-1}$ is of order $\Gamma$ for the fastest
moving material which dominates the emission. Thus
the brightness temperature in the frame of the 
synchrotron-emitting plasma is
$$
      T_B'=8\times 10^{11} \Gamma^{-3}\beta^{-2}
	  S({\rm mJy}) (\lambda/6\,{\rm cm})^2 t_d^{-2}d_{38}^2 
		\,\,{\rm K}
\eqno (3)
$$

On day 4, $S$= 9.9 mJy at 6-cm and from Equation (4) we obtain
$T_B'=5\times 10^{11}$ K.  If we insist that $T_B'<T_{\rm icc}$ then
we obtain $\Gamma^3\beta^2 \simgt 1$.  Thus we conclude that the radio
emitting region arises from a very fast moving shock, $\Gamma > 1.4$
or $\beta > 0.7$.  We now compute the total energy of this fast moving
radio emitting shock.  The energy is in the form of relativistic
electrons ($U_e$) and magnetic field strengths ($U_B$); there could be
a substantial amount of energy locked in protons but since protons do
not radiate we cannot constrain this additional energy.  It can be
shown that $U_B \propto S_p^{-4}\nu_p^{10}r^{11}$ and $U_e \propto
S_p^4 \nu_p^{-7} r^{-6}$ where $r$ is the radius of the source. The
total energy $U = U_e+U_B$ thus depends sensitively on
$r$. Furthermore, one term ($U_B$) increases rapidly with $r$ whereas
the other ($U_e$) decreases rapidly with $r$.  Unfortunately, we have
only a lower bound on $r$.

Scott \&\ Readhead\refto{Scott77} were faced with same dilemma when
investigating compact radio sources at meter wavelengths with 
ill-determined angular sizes. They noted that the total energy is
minimized by a radius obtained by setting $dU/dr=0$.
This radius also results in equipartition of
energy between the electrons and the magnetic field. Consequently,
this radius is referred to as the ``equipartition'' radius, $r_{eq}$ and
the resulting total energy as the ``equipartition'' energy, $U_{eq}$. 
The
equipartion angular {\it radius}, $\theta_{eq}=r_{eq}/d$ is\refto{Scott77} 
(note that many papers including ref.~\Ref{Scott77} define $\theta_{eq}$
to be a diameter and $S_\nu\propto\nu^{-\alpha}$, 
while we define $\theta_{eq}$ to be radius and $S_\nu\propto \nu^\alpha$)
$$
	\theta_{eq} = 
	120 
	\Bigl( {d   \over {\rm Mpc}} \Bigr)^{-1/17}
	\Bigl( {S_p \over {\rm mJy}} \Bigr)^{8/17}
	\Bigl( {\nu_p \over {\rm GHz}} \Bigr)^{(-2\alpha-35)/34}
	\,\,\mu{\rm arcsec}
	\eqno (4)
$$
It is convenient to express the total energy in terms of 
the equipartion energy,
$$
	{U\over U_{eq}} = {1\over 2}\eta^{11}(1+\eta^{-17})
	\eqno (5)
$$
where $\eta=\theta/\theta_{eq}$.

\refis{Scott77} 
	Scott, M. A. \&\ Readhead, A. C. S.
	The low-frequency structure of powerful radio sources
	and limits to departures from equipartition.
	{\it Mon. Not. Roy. Astr. Soc.} {\bf 180}, 539-550 (1977).

On day 16 $\nu_p=6$ GHz, $S_p=40$ mJy, so $\theta_{eq}=100\,\mu$arcsec
($r=6\times 10^{16}\,\rm cm$ at the distance of 38~Mpc), $B\sim
0.4\,\rm G$, and $U_{eq}=10^{49}$ erg.  Since $r/t=1.3c$, this
requires expansion of the radio photosphere at relativistic speed.
This equipartition angular size corresponds to a brightness
temperature $T_B=5\times 10^{10}\,\rm K$, well below $T_{\rm icc}$;
this $T_b$ is nothing but the ``equipartion'' temperature ($T_{eq}$)
of Readhead\refto{Readhead94}.  The electrons radiating at $\nu_p$
have energies of $\sim 60\,\rm MeV$.  If we had insisted on the size
determined by the fastest optical lines ($60,000\,\hbox{km s}^{-1}$),
the radius would be 7 times smaller, and therefore the source energy
would be dominated by relativistic electrons,
$U_e=10^{54}\,\hbox{erg}$ ---much larger than the total energy release
in a supernova.  Thus in addition to violating the X-ray constraint as
discussed in the previous section, this low velocity solution is thus
unacceptable on energetic grounds as well. 
This conclusion is
robust since earlier we arrived at the same conclusion using an 
entirely independent method, viz. the  
lack of variability and interstellar scattering.

If we require only that the source not violate the X-ray inverse
Compton constraint (Section 7), the brightness temperature could be
ten times the equipartition value, the radius thus 1/3 the
equipartition size (requiring $v=0.5c$), and $U_e=10^{52}\,\rm erg$.
We thus conclude that, on day 16, the combined energy in the electrons
and the magnetic field much lie in the range $10^{49}\,\hbox{erg}<U_e<
10^{52}\,\hbox{erg}$.

We can repeat this calculation for other dates. On days 4, 29 and 60,
we adopt $(\nu_p/\hbox{GHz},S_p/\hbox{mJy})=(10,15)$, (3,30) and (1,20)
respectively.
These give $\theta_{eq}=35, 160, 400$ $\mu$arcsec, and thus respectively
$r/t=1.9c$, $1.2c$ and $1.5c$. $U_{eq}=10^{48}\,\hbox{erg}$, $2\times 10^{49}
\,\hbox{erg}$ and $1.5\times 10^{49}\,\hbox{erg}$.
The angular sizes (and hence speeds) predicted above are verifiable
with VLBI observations.  The same observations will also directly
yield the brightness temperature and thus the actual particle and
field energies.

We close this discussion by noting some caveats.
\item{{\it (i).}} {\it External Absorption.}  We have ignored
any possible external (i.e. free-free) absorption in our analysis.  There
will be some level of free-free absorption from ionized circumstellar
material due to the progenitor star's wind.  It is clear from
the asymptotic convergence of the band-to-band spectral indicies to
$-1$ that least at late times there is little free-free absorption.
However, we expect the circumstellar material to be have a $r^{-2}$
distribution and thus free-free absorption could well be important at
early times. If so, the minimum energy and $\Gamma$ deduced above are
strict lower bounds.
\item{{\it (ii).}} {\it Peak Frequency.} The highest inferred
$\Gamma$ comes from the earliest observations, day 4.  Unfortunately,
on that day we observed the source in only two bands. This limited
coverage does not permit us to identify the true peak frequency.
However, for a simple self-absorbed synchrotron source, one can quite
easily demonstrate that inferring $T_B$ or $\theta_{eq}$ using flux
measurement at a frequency other than $\nu_p$ results in lower bounds
on both quantities.
\item{{\it (iii).}} {\it Expanding Synchrotron Source.} This, by far is
the most serious concern. Our simple analysis was done under the
framework of static synchrotron spheres.  However, the radio
photosphere is expanding at relativistic speeds.  The portion
expanding towards the observer becomes optically thin well before the
back portion. Thus the observed spectrum will be more complex than
that of a simple static synchrotron emitting sphere.  Indeed, there is
a suggestion that the broad-band spectrum on day 10 is not a pure
synchrotron self-absorbed spectrum. The flux in the 2-mm band (JCMT
observations) are well above a $\nu^{-1}$ extrapolation.  We are
engaged in a detailed analysis of this problem.

\noindent{\bf Geometry and Energetics}

Above we concluded that the radio emission in RSN 1998bw arises in a
shock with an initial $\Gamma\simgt 2$, which may have slowed to
transrelativistic speed by the end of our observing period.
The minimum total energy in the radio-emitting region
is $\sim 10^{49}$ erg, a rather impressive energy considering the fact
that most of the ejected mass is in a slower moving shock which manifests 
itself via the optical emission.

In the foregoing, we assumed spherical geometry. However, it could be
that the radio emission is in the form of jets.  This geometry has the
advantage of reducing the energy requirements considerably.
Fortunately, polarization measurements offer the possibility of probing
the geometry of the emitting region. However, as can be noted from
Table 1, we find no evidence for either circularly polarized signal or
linearly polarized signal.

The circular polarization of synchrotron radiation from relativistic
particles with a smooth pitch-angle distribution is small unless the
radiation is close to the cyclotron limit.
The absence of
circular polarization is thus not surprising, though it does place a
constraint on very slow jet models with magnetic fields well above
equipartition.

Shear and expansion in non-spherical synchrotron sources like jets
generally create magnetic fields with enough anisotropy that
their integrated polarizations are high.  For example
the integrated
fractional polarizations of radio quasars (including
the core, the jet and the lobes) at 5 and 8 GHz are rarely below 4\%
(ref. \Ref{Conway74}), and bursts in BL Lac objects (believed to be
caused by a single shock, as in the case of our supernova) have
integrated linear polarizations $\sim 10$\%\refto{Hughes89a,Hughes89b}.
The lack of any linearly polarized signal could be reconciled with the
jet hypothesis if the jet made an angle much
greater than $\Gamma^{-1}$ with respect to our line of sight.
Because of aberration, we would then
be seeing the back of the shock almost face-on,
so a tangled field in the shock plane would produce no net polarization.
However, this
hypothesis would require that the intrinsic $\Gamma$ and luminosity
of the jet are much larger than the ones inferred above.  This
seems implausible.

\refis{Conway74}
	Conway, R.G., Haves, P., Kronberg, P.P., Stannard, D., Vall\'ee,
	J.P., \&\ Wardle, J.F.C.
	The radio polarization of quasars.
	{\it M.N.R.A.S.} {\bf 168}, 137-162 (1974).

\refis{Hughes89a}
        Hughes, P.A., Aller, H.D. \&\ Aller, M.F.
        Synchrotron emission from shocked relativistic jets. I. The
        theory of radio-wavelength variability and its relation to superluminal
        motion.
        {\it Astrophys. J.} {\bf 341}, 54-67 (1989).

\refis{Hughes89b}
        Hughes, P.A., Aller, H.D. \&\ Aller, M.F.
        Synchrotron emission from shocked relativistic jets. I. A model
        for the centimeter wave band quiescent and burst emission from 
        BL Lacertae.
        {\it Astrophys. J.} {\bf 341}, 68-79 (1989).

The simplest interpretation of the observed low polarization is that we are
seeing a spherical blast wave. A VLBI image could
settle this uncertainty as well.

\noindent{\bf A new type of GRB?}

In our introduction we noted several circumstantial reasons favoring
the association of GRB 980425 to SNe 1998bw. Clearly, if this
association is correct then GRB 980425 is a new type of GRB, very
different and much less energetic
than the two GRBs at high redshift\refto{met97,srk98}.  Differing
in this conclusion are Wang \&\ Wheeler\refto{Wang98} who suggest that
all GRBs are associated with a SNe. In their model, the $\gamma$-ray
emission has two components, a low-brightness  isotropic component and
a  super-bright highly beamed component.  They advocate that the
$\gamma$-ray emission  of GRB 980425 arises from the isotropic
component and the gamma-ray emission from GRBs known to be at cosmological
distances is due to the beamed component.  We argue
this hypothesis is incorrect. The afterglow, especially the very
long-lived radio afterglow of cosmological GRBs such as that of GRB
970508 (ref. \Ref{Frail97}), is not subject to significant beaming.
Nonetheless, the radio afterglow of GRB 970508 is far more energetic
than the late time radio emission of SNe 1998bw. We conclude that if
GRB 980425 is indeed associated with SNe 1998bw then it is a member of
a new class of GRBs\refto{Bloom98b}.

Much theoretical effort \refto{Piran97, Meszaros97,Vietri97,Waxman97}
has been devoted to both the gamma-ray emission and the subsequent
afterglow (which was indeed predicted before it was observed) of
cosmological GRBs.  The models give a reasonable description of the
observed phenomena thereby boosting our confidence in them.  The
models posit that a large amount of energy $E_0$ is mysteriously
supplied to an astonishingly small mass of ejecta, $M_{ej}\sim 10^{-5}
M_\odot$. The result is a very high-$\Gamma$ shock.


After day 4, the synchrotron lifetimes of the electrons in SN 1998bw
radiating at $\sim 5$ GHz are (for equipartition fields) $\sim 1\,{\rm yr}$.
[Note that the total energy radiated in the observed part of the
radio spectrum is $\sim 10^{45}\,\hbox{erg}$, much less than $U_{eq}$.
But the flat spectrum implied by the flux in the 2-mm band 
suggests that higher energy electrons may be more nearly radiative].
The X-ray flux limit implies that the inverse Compton losses cannot
be much greater, so the shock appears to be non-radiative throughout the
observing period.  Extrapolating back in time in a simple equipartition
model in which  pressure $p\propto r^{-2}$ and the particle spectrum
is independent of radius, the optically thin synchrotron luminosity
from radius $r$ to $2r$ scales as $r^{-1}$, as does the ratio of
first inverse Compton to synchrotron luminosity.  Thus at early
times a shock in the circumstellar gas or an internal shock in the
ejecta would be radiative and predominantly a gamma-ray source.

It is of some interest to note that more than two decades ago,
Colgate\refto{Colgate74} had proposed gamma-ray emission from SNe as
possible origin of GRBs. The proposed mechanism, acceleration of the
supernova shock to relativistic speed down the density gradient of a
massive stellar envelope, has not been confirmed by more detailed
calculations including radiation coupling (e.g. ref.~\Ref{Ensman92}).
Yet it might work with the much smaller and less massive
envelopes of Carbon and Helium stars appropriate to type Ib/c
supernovae.  The gamma-ray burst energy is $10^{48}$ erg which is
smaller than the inferred minimum energy in the relativistic shock.
Thus on energetic grounds, the relativistic shock can easily account
for the observed gamma-rays. The outstanding theoretical issue now is
how to generate a shock the required $\Gamma$ and total energy, and
avoid thermalizing the emitted spectrum.

\refis{Ensman92}
	Ensman, L. \&\ Burrows, A.
	Shock Breakout in SN 1987A.
	{\it Astrophys. J.} {\bf 393}, 742-755 (1992).

\refis{Colgate74}
	Colgate, S. A. 
	Early gamma rays from supernovae. 
	{\it Astrophys. J.} {\bf 187}, 333-335 (1974).

Having established a physically motivated connection between SN 1998bw
and GRB 980425 we now consider other such potential associations.  In
our view, the key factor which might make possible $\gamma$-ray bursts from
SNe is the relativistic shock. At radio wavelengths, as argued
extensively in this paper, the relativistic shock is manifested by high
brightness temperature. Shklovskii\refto{Shklovskii85} and
later Slysh\refto{Slysh90} noted that type Ib/Ic RSNe generally
exhibit higher $T_B$ than type II RSNe.

\refis{Slysh90}
        Slysh, V.  I. 
        Synchrotron self-absorption of radio emission from
        supernovae.
        {\it Sov. Astron. Lett.} {\bf 16}, 339-342 (1990).

The earlier studies\refto{Shklovskii85,Slysh90,Chevalier98} noted the
the possible presence of high velocity gas in type Ib/c RSNe.
However, as reviewed elsewhere\refto{Kulkarni98}, these earlier
studies may have failed to appreciate how robust the arguments for
high speeds were.  If we insist that only a small portion of the SNe
energy release should go into the radio-emitting shocks then we are
forced to conclude that type Ib/Ic supernovae have shocks of at least
transrelativistic speed.

As noted elsewhere\refto{Kulkarni98}, the earliest measurements offer
the greatest diagnostic of high-$\Gamma$ shocks. For example, for SN
1983N, the very first radio detection was 2 mJy in the 6-cm band
(ref. \Ref{Weiler86}).  Following Chevalier\refto{Chevalier98}), the
assumed epoch of this measurement is day 1 and a distance is 5.4
Mpc. The inferred $T_b' \sim T_{eq}$, requiring $\Gamma^3\beta^2
\simgt 1$ or $\Gamma \simgt 1.3$.  Slysh\refto{Slysh90} and Chevalier
used later measurements and consequently missed the high speed
shock. Unfortunately prompt radio observations of SNe are rare.  Thus
it is possible that relativistic shocks exist in other type Ib/Ic RSNe
and were missed through lack of suitable observations.

We presented radio observations of SN 1998bw, a type Ic SN which has
been associated with GRB 980425 on purely probabilistic grounds.  This
is the most luminous radio SN to date.  Assuming that the radiation
is synchrotron emission, we conclude that the shock in this SN must
have a speed close to that of light, and a minimum energy $\sim
10^{49}\,\rm erg$.  This relativistic shock has apparent speed of 2$c$
on day 4, slowing down to $c$ a month later.  By analogy with GRBs we
suggest that this relativistic shock could have generated a burst of
gamma-rays at very early times. Thus our work provides a direct
physical link between GRB 980425 and SN 1998bw.  If this
identification is correct, the variable NFI X-ray source
1SAXJ1935.3-5252 must have been an unrelated object, more likely to be
a variable AGN than a true transient.  It should therefore reappear in
future X-ray observations.

As with any advance, there are many significant
open questions.  Is the shock spherical or collimated?  How much energy
beyond the minimum
is involved in the relativistic shock? Is this phenomenon common to all
Type Ic SNe? Fortunately, these questions can be answered by
observations. Early time observations provide the best diagnostic
of the fastest shock. VLBI observations could directly measure the geometry
of the fast shock. And finally, it would be worth following up radio
observations of GRBs with profiles similar to that of GRB 980425
(ref. \Ref{Bloom98b}).

\refis{Bloom98b}
	Bloom, J. S.  Kulkarni, S.R., Harrison, F., Prince, T., \&\
	Frail, D.A.
	Type Ib/Ic supernovae: A new class of gamma-ray bursts?
	in preparation, (1998).


\noindent{\bf Acknowledgments.}
We gratefully acknowledge Lorne Avery and Gerald
Moriarty-Schieven for their help in making the JCMT observations.  
DAF thanks M. Rupen for useful discussions.  
SRK thanks
A. Readhead for extensive discussions of brightness
temperature.  The
Australia Telescope is funded by the Commonwealth of Australia for
operation as a National Facility managed by CSIRO.  The James Clerk
Maxwell Telescope is operated by The Joint Astronomy Centre on behalf
of the Particle Physics and Astronomy Research Council of the United
Kingdom, the Netherlands Organization for Scientific Research, and the
National Research Council of Canada.  The VLA is a facility of the
National Science Foundation operated under cooperative agreement by
Associated Universities, Inc.  The research of SRK and of ESP
is supported by the
National Science Foundation and NASA.


\bigskip
\centerline {\bf References}
\bigskip
\endreferences

\vfill\eject


\topinsert{
$$
\table
\tablespec{\l\l\l\l\l\l\l\l}
\body{
\header{\bf Table 1. Radio Flux Densities Measurements of SN 1998bw$^{(a)}$}
\skip{15pt}
\hline
\skip{5pt}
Date & Elapsed & S$_{20}$ & S$_{13}$ & S$_{6}$ & S$_{3}$ & Array & Int
Time \end
(UT) & Time~(days) & (mJy) & (mJy) & (mJy) & (mJy)& Config. & (hrs) \end
\skip{5pt}
\hline
\skip{5pt}
1998 Apr 28 & $\phantom{1}$3.0    & \omit & \omit &  9.0  & 13.0 &
750A & 3 \end
1998 Apr 29 & $\phantom{1}$4.0    & \omit & \omit &  9.9  & 13.0 &
750A & 2 \end
1998 May 05 & $\phantom{1}$9.9    & \omit & \omit & 39.0  & 48.0 &
750A & 9 \end
1998 May 07 & 11.7   & 6.2   & 19.7  & 44.6  & 49.4 & 750A & 3 \end
1998 May 10 & 14.6   & 7.7   & 22.3  & 39.9  & 37.6 & 750A & 2 \end
1998 May 11 & 15.7   & 9.2   & 23.5  & 37.4  & 34.3 & 750A & 2 \end
1998 May 12 & 16.5   & 8.9   & 23.9  & 37.1  & 31.4 & 750A & 1 \end
1998 May 13 & 17.8   & 11.0  & 25.1  & 32.6  & 26.2 & 6C & 3   \end
1998 May 15 & 19.7   & 12.1  & 25.3  & 28.6  & 21.6 & 6C & 0.5 \end
1998 May 17 & 21.6   & 12.7  & 20.9  & 24.3  & 18.8 & 6C & 1   \end
1998 May 19 & 23.6   & 11.8  & 22.9  & 24.7  & 17.6 & 6C & 1   \end
1998 May 21 & 25.9   & 16.7  & 28.0  & 27.6  & 20.9 & 6C & 0.5 \end
1998 May 22 & 26.8   & 15.8  & 28.7  & 29.5  & 21.7 & 6C & 0.7 \end
1998 May 24 & 28.8   & 19.6  & 31.1  & 30.0  & 22.0 & 6C & 1.3 \end
1998 May 25 & 30.0   & 20.0  & 31.3  & 30.0  & 22.1 & 6C & 2   \end 
1998 May 28 & 32.9   & 23.7  & 27.3  & 30.3  & 21.3 & 750E & 1 \end
1998 May 30 & 34.7   & 23.9  & 33.5  & 28.6  & 20.2 & 750E & 1 \end 
1998 Jun 01 & 36.8   & 23.5  & 31.8  & 27.0  & 18.4 & 750E & 1 \end
1998 Jun 03 & 38.8   & 25.2  & 31.0  & 24.6  & 16.1 & 750E & 0.8 \end
1998 Jun 04 & 40.0   & 25.9  & 31.3  & 24.1  & 16.6 & 750E & 0.6 \end
1998 Jun 10 & 45.7   & 28.9  & 26.8  & 20.7  & 13.2 & 750E & 1 \end
1998 Jun 16 & 51.7   & 25.8  & 23.1  & 16.3  & 10.5 & 750E & 2 \end
1998 Jun 22 & 57.7   & 19.7  & 18.5  & 14.0  & 8.1 & 750E & 2 \end
\skip{5pt}
\hline
}
\endtable
$$
}\endinsert

\par\noindent
(a) The entries (from left to right), the UT date of the observation,
the time in days since the start of the SN explosion (calculated
assuming that it occurred when the gamma-rays from GRB 980425 were
first detected on 1998 March 25.90915 UT (ref. \Ref{Soffitta98})), the flux
density in milliJy at 20 cm (1.38 GHz), 13 cm (2.49 GHz), 6 cm
(4.80 GHz), and 3 cm (8.64 GHz), the array configuration for the ATCA,
and the total integration time obtained for each pair of wavelengths
(20 cm/13 cm and 6 cm/3 cm). 

\noindent
(b) All observations used a bandwidth of 128 MHz and two orthogonal
linear polarizations for each wavelength pair. The individual antenna
elements were moved in the course of this monitoring effort, forming
different array configurations (750A, 6C, 750E).  
We minimized the effects of confusion, an important issue for compact
configurations such as 750A and 750E, by subtracting background
sources from the visibilities, and by excluding the shortest baselines
from the analysis.

\noindent 
(c) The initial pointing centre was toward \1SAXJ1935\ but on 1998 May
07 it was shifted to SN 1998bw.  All the tabulated flux density
measurements of SN 1998bw prior to this date were corrected by the
primary beam response of the antennas.

\noindent
(d)
A search was made for linear or circularly polarized flux on day
12. No signal was detected in the Stokes Q, U and V above 0.5\% and
0.9\% at 3 cm and 6 cm, respectively. The limits on day 29 and 30
are $<1.5\%$ polarization at 3, 6 and 13 cm and $<2\%$ linear
polarization at 20 cm.

\vfill\eject

\centerline{{\psfig{file=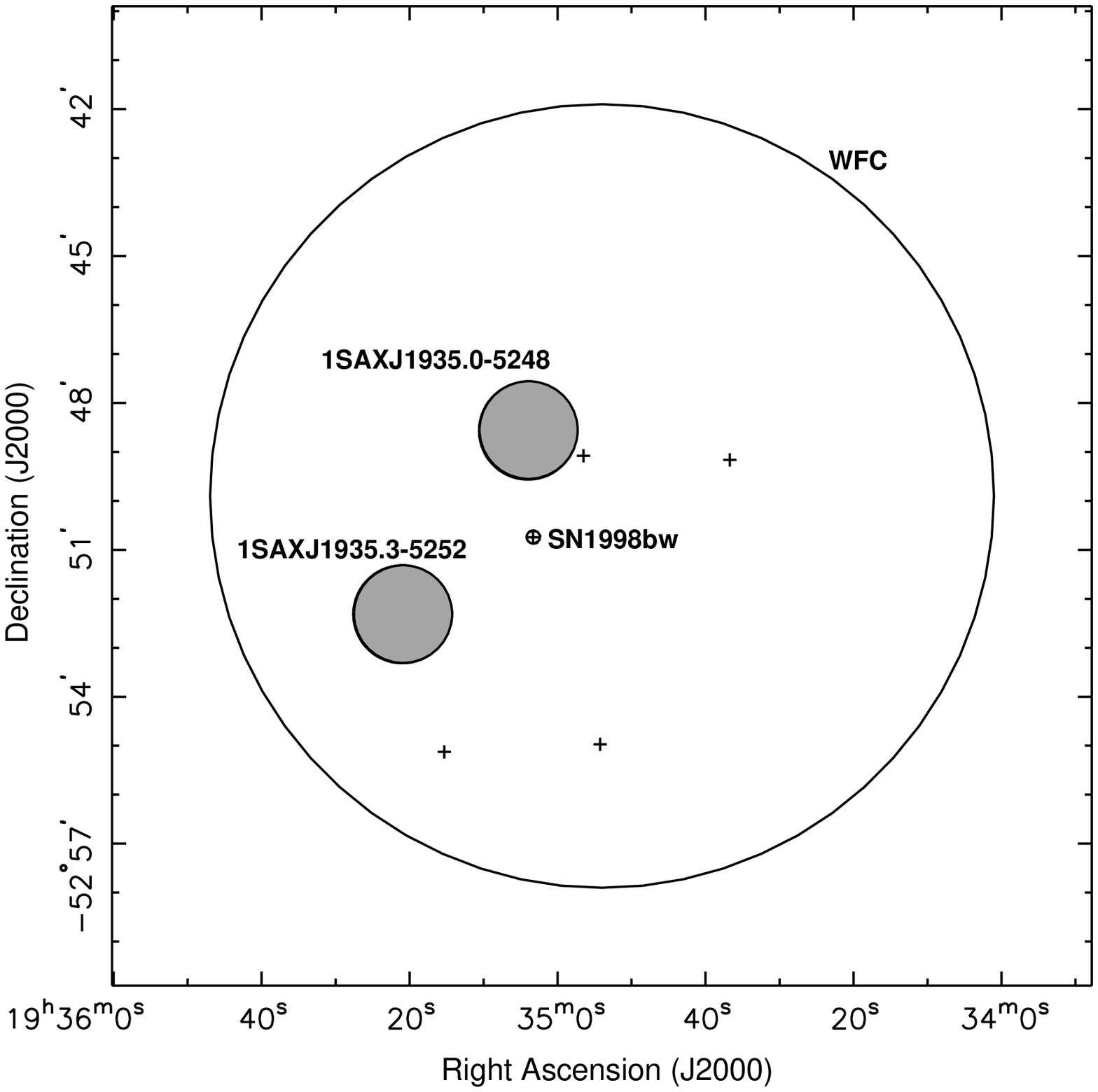,width=5.0in,clip=0,angle=0}}}
\noindent Figure \localization. The position of objects within the
8-arcmin (radius) location error circle of the WFC for GRB 980425
(ref. \Ref{Soffitta98}).  The two NFI X-rays sources\refto{Pian98g}
(the transient 1SAXJ1935.3$-$5252 and the steady
1SAXJ1935.0$-$5248) are indicated by hatched
regions indicating their approximate position uncertainties. Small
crosses indicate the positions of radio sources detected by the ATCA.
The radio and optical emission from SN 1998bw is given by a circle with
a cross.  The mean position for SN 1998bw, obtained by averaging the
best 3 cm and 6 cm observations, is $\alpha=$ 19$^{\rm h}$35$^{\rm
m}$3\rlap{.}{$^{\rm s}$}316 and
$\delta=-$52$^\circ$50$^{\prime}$44\rlap{.}{$^{\prime\prime}$}75,
(equinox J2000).  The 1-$\sigma$ error in $\alpha$ is 0\rlap{.}{$^{\rm
s}$}01 and 0\rlap{.}{$^{\prime\prime}$}07 in $\delta$.

\vfill\eject

\centerline{{\psfig{file=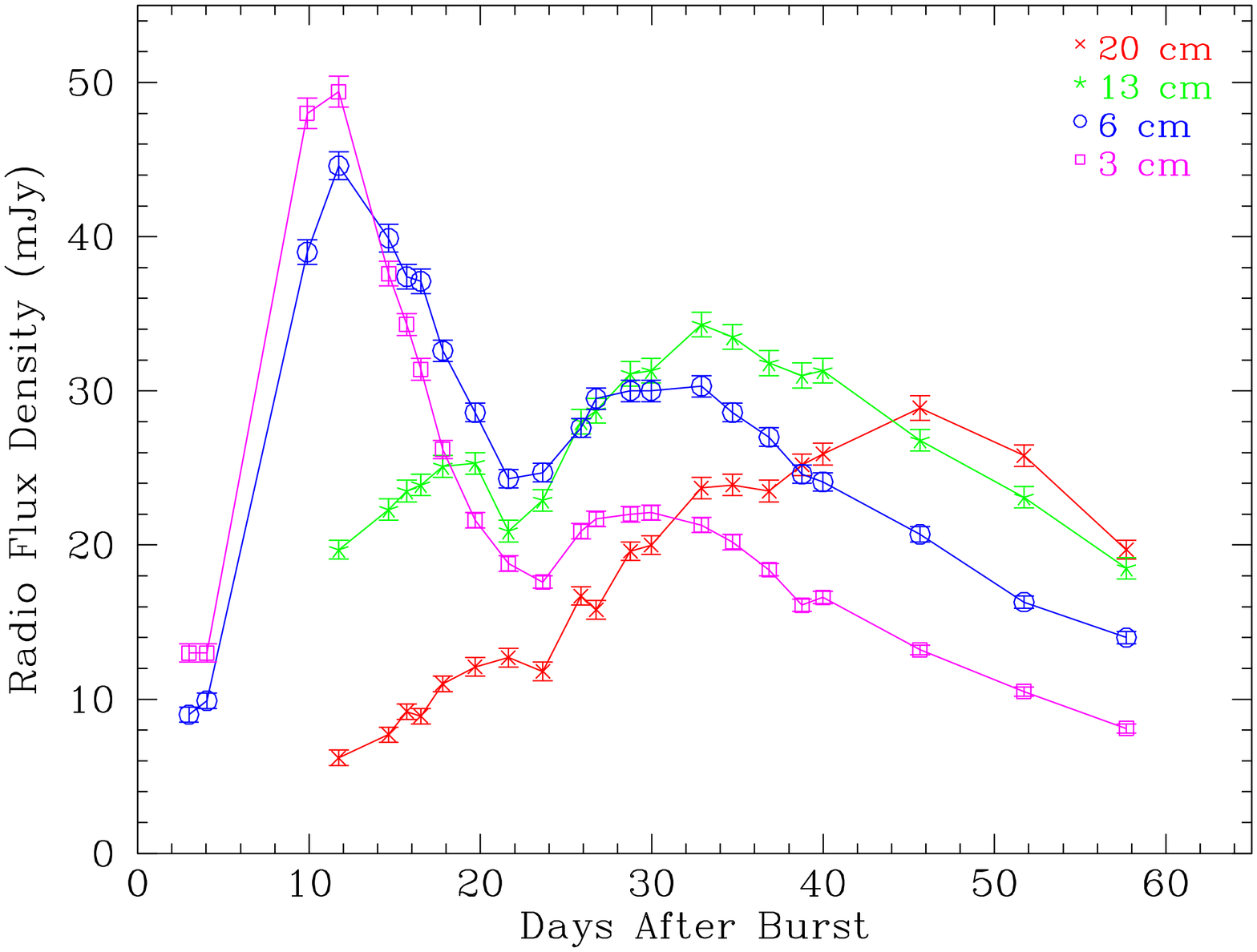,width=6.0in,clip=0,angle=270}}}
\noindent {\bf Figure \lightcurves.--} The radio light curve of SN
1998bw. Four wavelengths, 20 cm (cross), 13 cm (star), 6 cm (circle)
and 3 cm (square) are plotted together. The age of the supernova has
been calculated assuming that the explosion date can is given by the
detection of gamma-rays from \grb. The error bars given on the plot are
larger than the formal errors estimated from the receiver noise owing
to the difficulties of determining the flux density with short
snapshots.
The errors, including absolute flux scale errors, thermal noise and
confusion noise,   can be approximated by the quadrature sum of a
constant term (0.5 mJy for both 20 cm and 13 cm, 0.3 mJy for 6 cm and
0.2 mJy for 3 cm) and a term which is a fraction of the flux density
(2\%).

\vfill\eject

\centerline{\hbox{\psfig{file=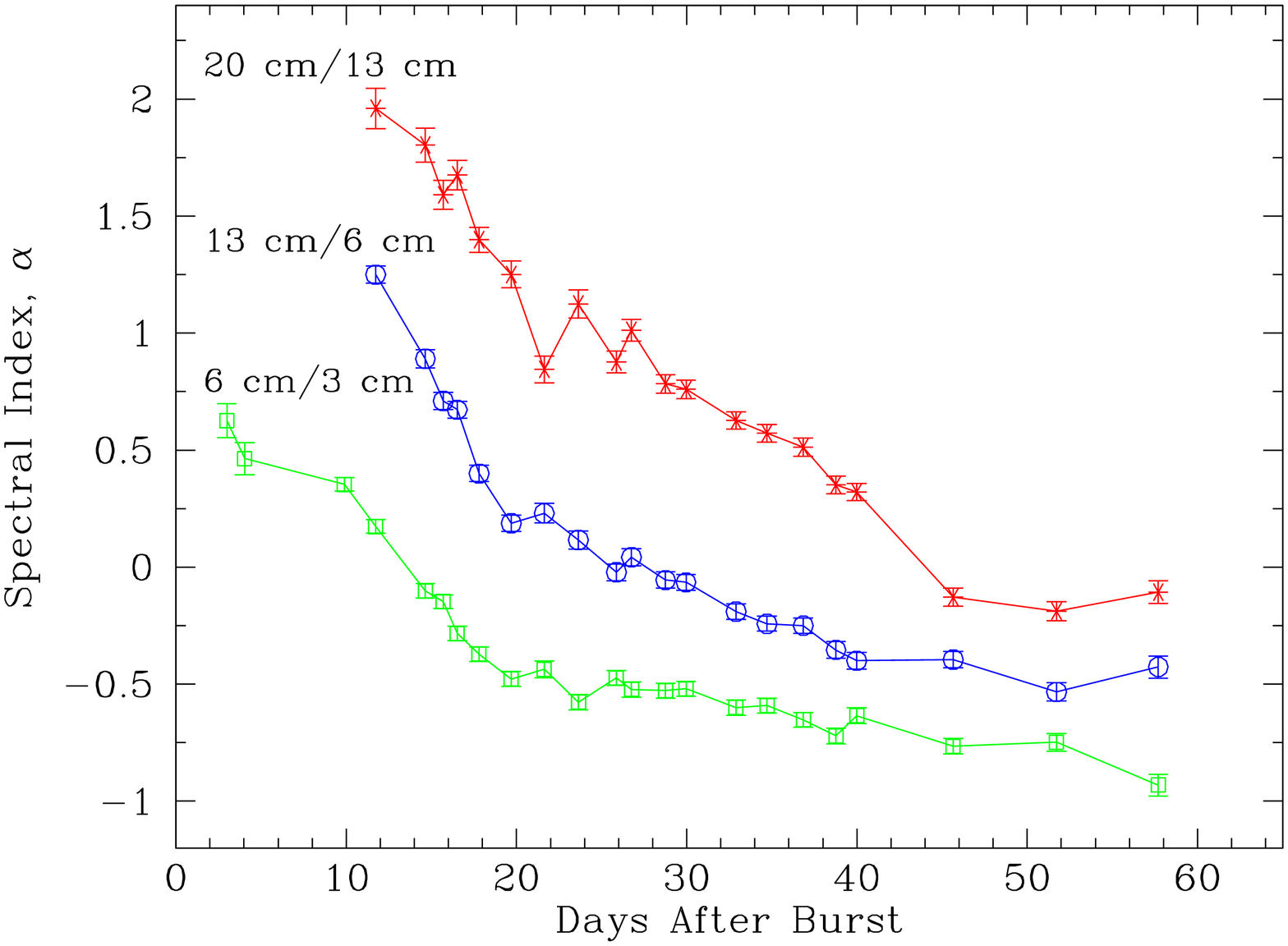,width=7.0in,clip=0,angle=270}}}
\noindent {\bf Figure \spectralindex.--} The evolution of the radio
spectral index for SN 1998bw. The spectral index $\alpha$ (where
S$_\nu\propto\nu^\alpha$) is calculated between 20 cm and 13 cm
(cross), 13 cm and 6 cm (open circle) and 6 cm and 3 cm (squares).  The
age of the supernova has been calculated assuming that the explosion
date is given by the detection of gamma-rays from \grb.

\vfill\eject

\centerline{\hbox{\psfig{file=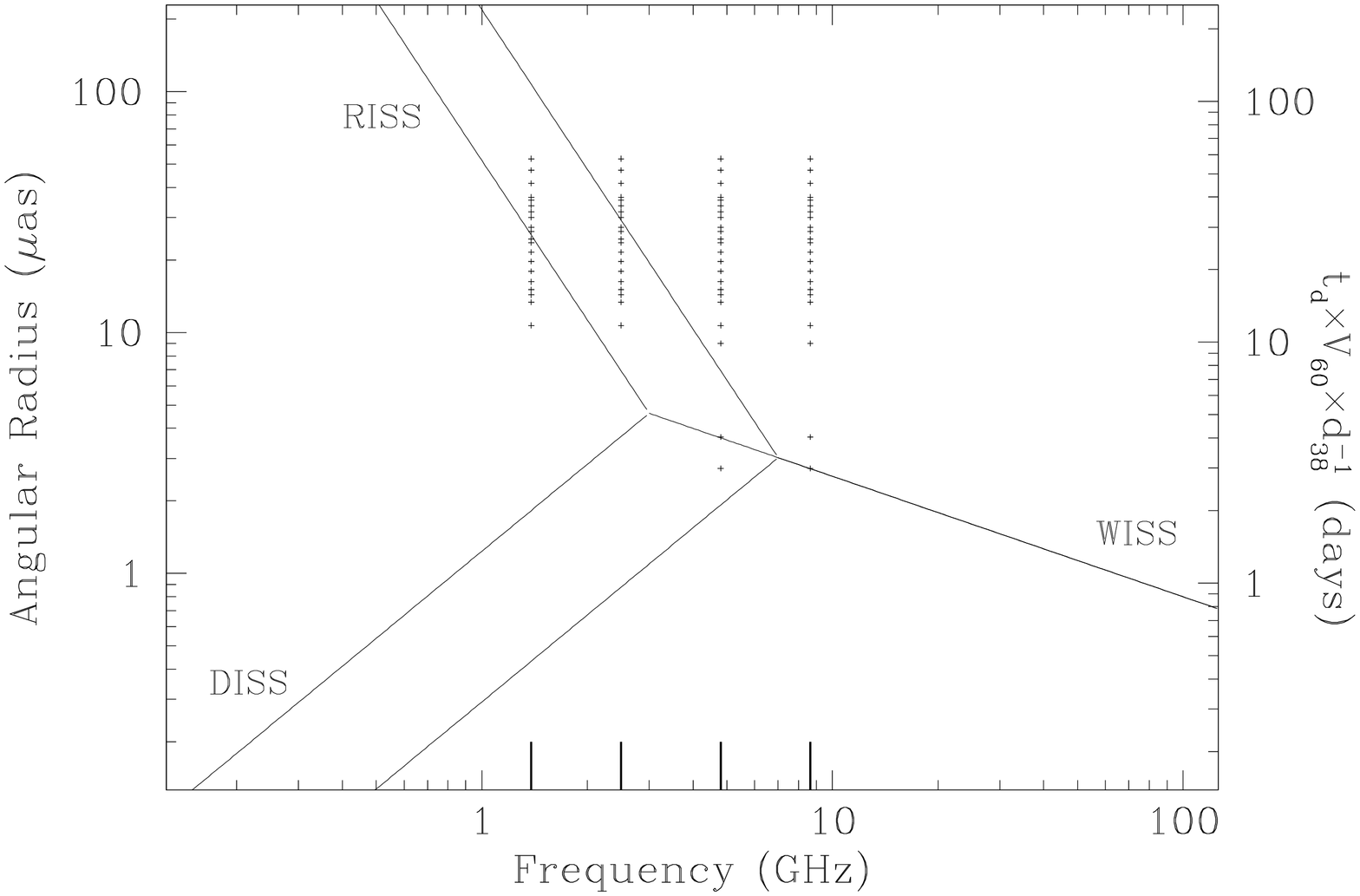,width=7.0in,clip=0,angle=270}}}
\noindent {\bf Figure \ISS.--} The different regimes of interstellar
scattering. Lines indicate the frequency dependence of diffractive
(DISS), refractive (RISS) and weak (WISS) scattering as a function of
the source angular radius (left hand axis). The exact value of the
transition frequency between strong and weak scattering is uncertain
($\nu_0$=3-7 GHz) and this is reflected in the plot by the different
lines. On the right hand axis we plot the expected time (in days) that
SN 1998bw would reach this radius if the velocity of the
radio photosphere in units of $60,000\,\hbox{km s}^{-1}$
$v_{60}$=1 and the source distance in units of 38~Mpc, d$_{38}$=1. Bold
ticks at the bottom of the figure indicate the frequencies used in the
ATCA observations. The small crosses are plotted at the frequencies and
dates at which ATCA observations were made.

\vfill\eject

\centerline{\hbox{\psfig{file=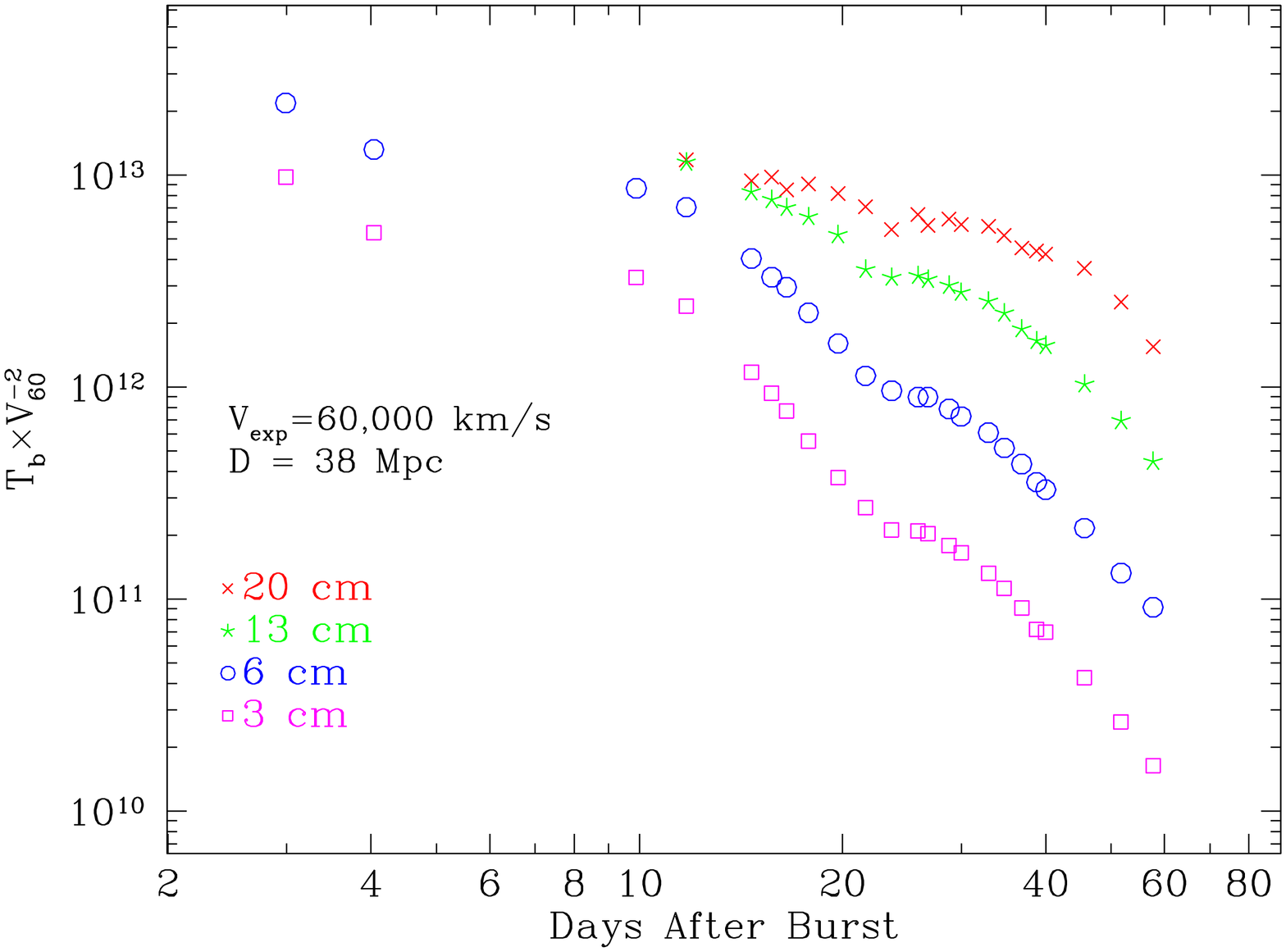,width=7.0in,clip=0,angle=270}}}
\noindent {\bf Figure \TB.--} The evolution of the brightness
temperature of SN 1998bw at 20, 13, 6 and 3 cm. 
The y-axis is the brightness temperature assuming a distance of 38 Mpc
and a velocity of $v=60,000v_{60}$ km s${-1}$.

\endmode
\bye